\documentclass[showkeys,showpacs,prd]{revtex4}
\usepackage{graphicx}
\usepackage{amsmath}
\usepackage{amsfonts}
\usepackage{color}

\begin{document}

\title{SU(3) glueball gluon condensate}
\author{Vladimir Dzhunushaliev}
\email{vdzhunus@krsu.edu.kg}
\affiliation{Institute for Basic Research,
Eurasian National University,
Astana, 010008, Kazakhstan; \\
Institut f\"ur Physik, Universit\"at Oldenburg, Postfach 2503
D-26111 Oldenburg, Germany}

\begin{abstract}
In a scalar approximation the distribution of a gluon condensate in a glueball is calculated. In this approximation the SU(3) gauge fields are separated on two parts: (1) is the $SU(2) \subset SU(3)$ subgroup, (2) is the coset 
$SU(3) / SU(2)$. Using an approximate nonperturbative quantization technique two scalar fields are applied for the description of the SU(2) and coset degrees of freedom. In this approach 2-point Green's functions are a bilinear combination of scalar fields and 4-point Green's functions are the product of 2-points Green's functions. 
\end{abstract}


\maketitle

\section{Introduction}

One of the problems in a nonperturbative QCD is the determination of  condensates. It is well known that the condensates can only be determined in a nonperturbative formulation of the QCD. There is a long history of attempts to determine the gluon condensate from first principles \cite{Banks}, \cite{Shifman:1978bx}.

Glueball is (thought to be) bound state of gluons. The nonlinear properties of gluons create the possibility of a color-neutral state made of gluons only: glueballs. Glueball properties cannot be computed with perturbation theory, and these remain very mysterious objects over thirty years after QCD was understood in its workings. For review, see Ref. \cite{Mathieu:2008me}. 

In Ref. \cite{Dzhunushaliev:2010qs} the distribution of a gluon condensate in a flux tube stretched between quark and antiquark is calculated using approximate nonperturbative quantization technique. It is shown that a longitudinal chromoelectric field is confined with a surrounding coset chromomagnetic field. Such picture presents the concrete realization of dual QCD model in a scalar model of the flux tube. In the scalar model the SU(3) gauge fields are separated on two parts: (1) is the $SU(2) \subset SU(3)$ subgroup, (2) is the coset $SU(3) / SU(2)$. The SU(2) degrees of freedom are almost classical and the coset degrees of freedom are quantum ones. A nonperturbative approach for the quantization of the coset degrees of freedom is applied. In this approach 2-point Green's functions are a bilinear combination of scalar fields and 4-point Green's functions are the product of 2-points Green's functions. The gluon condensate is an effective Lagrangian describing the SU(2) gauge field with broken gauge symmetry and coupling with the scalar field. Corresponding field equations give us the flux tube.

Here we continue the research in this direction and will calculate the distribution of gluon condensate in the glueball. In Ref. \cite{Dzhunushaliev:2003sq} the model of the glueball based on the calculations similar to \cite{Dzhunushaliev:2010qs}. Only one difference is that coset degrees of freedom are quantized as well. In this notice we would like to obtain the distribution of gluon condensate for the glueball using results from Ref. \cite{Dzhunushaliev:2003sq} and compare obtained results with the results of Ref. \cite{Dzhunushaliev:2010qs}. 

\section{Scalar model of glueball}

Following to \cite{Dzhunushaliev:2003sq} we separate SU(3) degrees of freedom on two parts: the first is the $SU(2) \subset SU(3)$ subgroup, the second is the coset $SU(3) / SU(2)$. We average SU(3) Lagrangian using some assumptions and approximations. The main idea in this approximation is that 2 and 4-points Green's functions can be approximately described by two scalar fields 
$\phi, \chi$: 
\begin{align}
	\left( G_2 \right)^{ab}_{\mu \nu}(x_1, x_2) & = 
	\left\langle
		A^a_\mu (x_1) A^b_\nu (x_2)
	\right\rangle 
	& \approx & \; 
	C^{ab}_{\mu \nu} \phi(x_1) \phi^*(x_2) +
	m^{ab}_{\mu \nu},
\label{1-10}\\
	\left( G_4 \right)^{abcd}_{\mu \nu \rho \sigma}(x_1, x_2, x_3, x_4) &= 
	\left\langle 
		A^a_\mu (x_1) A^b_\nu (x_2) A^c_\rho (x_3) A^d_\sigma (x_4)
	\right\rangle 
	&\approx & \; 
	\left\langle
		A^a_\mu (x_1) A^b_\nu (x_2)
	\right\rangle
	\left\langle
		A^c_\rho (x_3) A^d_\sigma (x_4)
	\right\rangle
\label{1-20}\\
	\left( G_2 \right)^{mn}_{\mu \nu}(x_1, x_2) &= 
	\left\langle
		A^m_\mu (x_1) A^n_\nu (x_2)
	\right\rangle 
	&\approx & \; 
	C^{mn}_{\mu \nu} \chi(x_1) \chi^*(x_2) +
	m^{mn}_{\mu \nu},
\label{1-30}\\
	\left( G_4 \right)^{mnpq}_{\mu \nu \rho \sigma}(x_1, x_2, x_3, x_4) &= 
	\left\langle
		A^m_\mu (x_1) A^n_\nu (x_2) A^p_\rho (x_3) A^q_\sigma (x_4)
	\right\rangle 
	&\approx & \; 
	\left\langle
		A^m_\mu (x_1) A^n_\nu (x_2)
	\right\rangle
	\left\langle
		A^p_\rho (x_3) A^q_\sigma (x_4)
	\right\rangle
\label{1-40}
\end{align}
where $a,b,c,d = 1,2,3$ are SU(2) indexes, $m,n,p,q = 4,5,6,7,8$ are coset indexes and $C^(\cdots)_(\cdots)$ and $m^(\cdots)_(\cdots)$ are some constants \footnote{Let us note that in \cite{Dzhunushaliev:2003sq} we used more than one fields: $\phi^a$ instead of $\phi$ and $\phi^m$ instead of $\chi$. This distinction is unessential. We use fields $\phi, \chi$ in order to be close as soon as it is possible to the designation of the paper \cite{Dzhunushaliev:2010qs}}. After that the effective Lagrangian is
\begin{equation}
	\mathcal L_{eff} = \left\langle \mathcal L_{SU(3)} \right\rangle = 
	\frac{1}{2} \left| \nabla_\mu \phi \right|^2 -
	\frac{\lambda_1}{4} \left(
		\left| \phi \right|^2 - \phi_\infty^2
	\right)^2 +
	\frac{1}{2} \left| \nabla_\mu \chi \right|^2 -
	\frac{\lambda_2}{4} \left(
		\left| \chi \right|^2 - \chi_\infty^2 
	\right)^2 + \frac{\lambda_2}{4} \chi^4_\infty - 
	\frac{1}{2} \phi^2 \chi^2 
\label{1-50}
\end{equation}
where $\lambda_{1,2}$ and $\phi_\infty, \chi_\infty$ are some parameters. We use here $(+,-,-,-)$ metric signature. It is necessary to emphasize: \textcolor{blue}{The effective Lagrangian \eqref{1-50} is some approximation for the nonperturbative quantization of the SU(3) gauge theory.}

The physical significance of this approach is following: 
\begin{itemize}
	\item The scalar fields $\phi, \chi$ describe nonperturbative quantized $SU(2)$ and coset $SU(3) / SU(2)$ degrees of freedom correspondingly. 
	\item The terms $\left| \nabla_\mu \phi \right|^2$ and 
	$\left| \nabla_\mu \chi \right|^2$ in the effective Lagrangian \eqref{1-50} appear after the nonperturbative quantum averaging of the terms 
	$( \nabla_\mu A^B_\nu )^2$ in initial SU(3) Lagrangian. 
	\item The terms $\phi^4, \chi^4$ appear after the nonperturbative quantum averaging of the term $f^{ABC} f^{AMN} A^B_\mu A^C_\nu A^{M \mu} A^{N \nu}$. 
	\item The term $\phi^2 \chi^2$  appears after the nonperturbative quantum averaging of the term $f^{Aab} f^{Amn} A^a_\mu A^b_\nu A^{m \mu} A^{n \nu}$.
	\item The terms $\phi^2 \phi^2_\infty$, $\phi^4_\infty$, 
	$\chi^2 \chi^2_\infty$ and $\chi^4_\infty$ are some additional assumptions. 
\end{itemize}
The field equations describing glueball in the scalar model are 
\begin{eqnarray}
  \partial_\mu \partial^\mu \phi &=& 
  - \phi \left[ \chi^2 + \lambda_1 
  \left(
    \phi^2 - \phi_\infty 
  \right) \right],
\label{1-60}\\
  \partial_\mu \partial^\mu \chi &=& 
  - \chi \left[ \phi^2 + \lambda_2 
  \left(
    \chi^2 - \chi_\infty 
  \right) \right].
\label{1-70}
\end{eqnarray}
Special feature of equations \eqref{1-60} \eqref{1-70} is that they have regular solutions with special choice of $\phi_\infty, \chi_\infty$ parameters. One can say that these equations are eigenvalue problem with 
$\phi_\infty, \chi_\infty$ eigenvalues. 

\begin{figure}[h]
  \begin{minipage}[t]{.45\linewidth}
  \begin{center}
    \includegraphics[width=.6\linewidth,angle=-90]{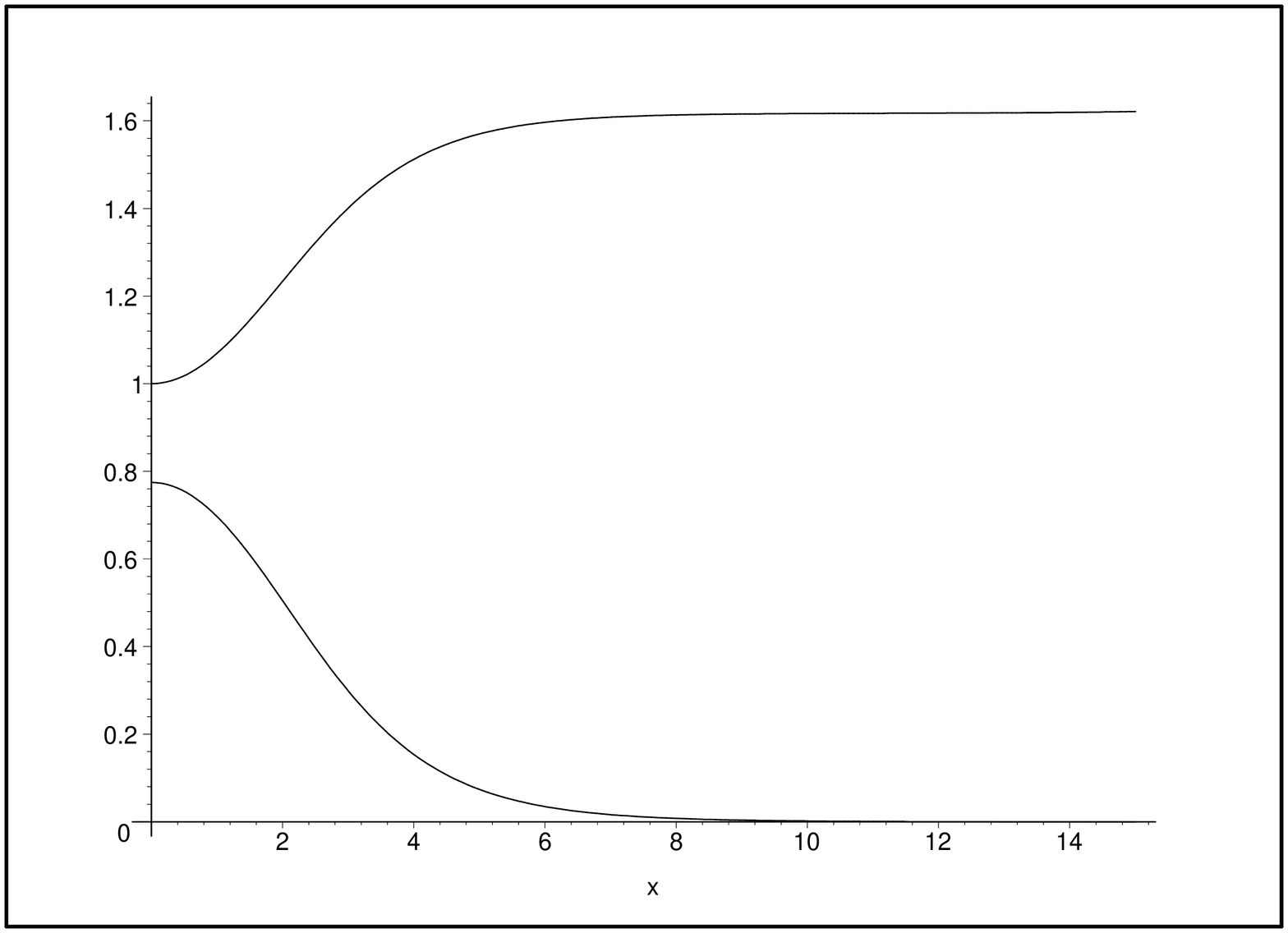}
    \caption{The profiles of $\phi(x)$ - upper curve, $\chi(x)$ - lower curve; 			$\lambda_1 = 0.1$, $\lambda_2 = 1.$, $\phi_\infty = 1.6171579$, 
    $\chi_\infty = 1.49273856$.}
    \label{functions}
  \end{center}  
  \end{minipage}\hfill
  \begin{minipage}[t]{.45\linewidth}
  \begin{center}
    \includegraphics[width=.6\linewidth,angle=-90]{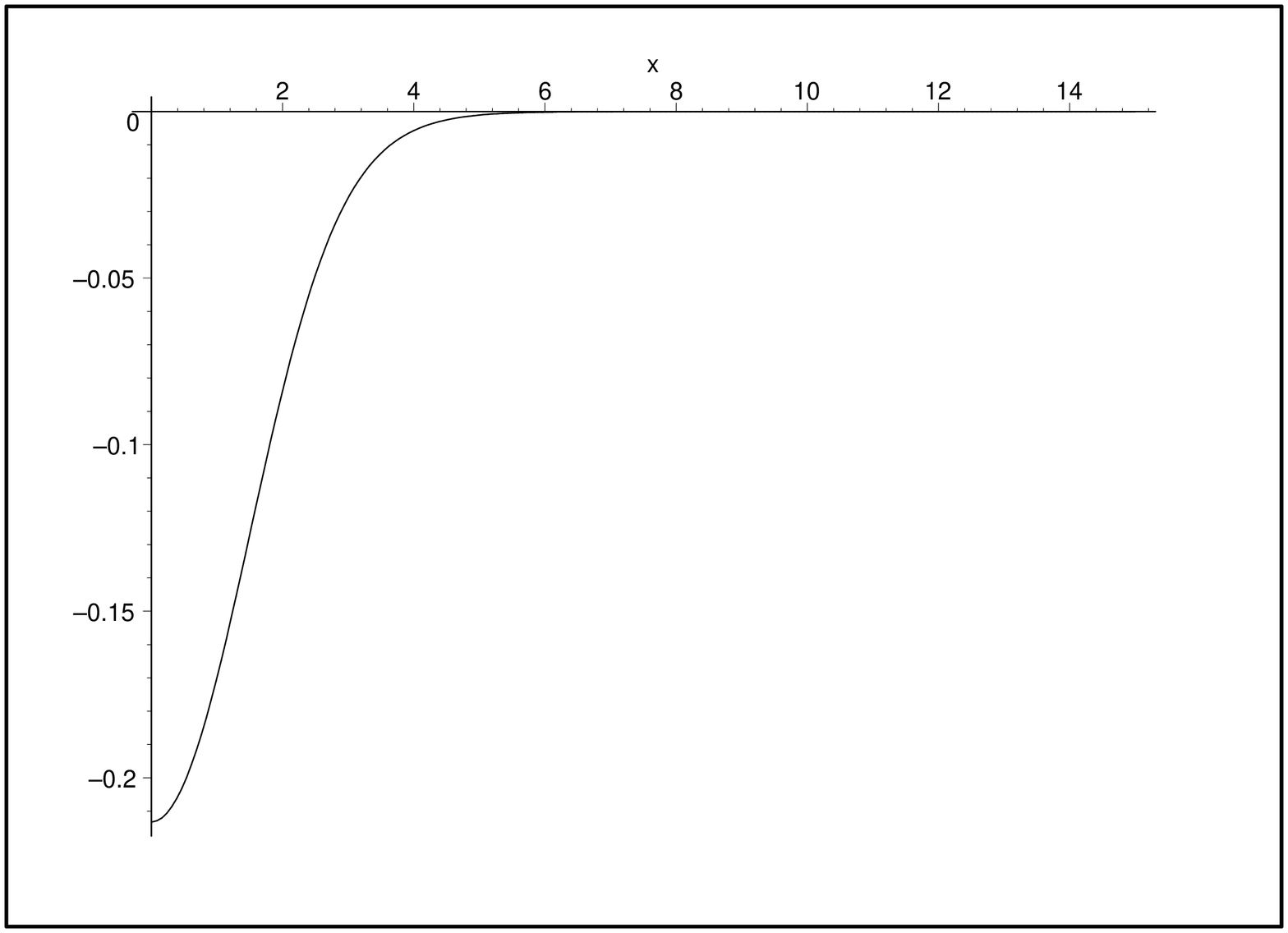}
    \caption{The profile of the SU(3) gluon condensate.}
    \label{condensate}
  \end{center}  
  \end{minipage} 
\end{figure}

A spherically symmetric regular solution of \eqref{1-60} \eqref{1-70} equations describes a ball filled with fluctuating quantum SU(3) gauge fields. \textcolor{blue}{We interpret this ball as the glueball.} In the spherically symmetric case equations \eqref{1-60} \eqref{1-70} have following form
\begin{eqnarray}
   \tilde \phi'' + \frac{2}{x} \tilde \phi' &=& \tilde \phi
    \left[
      \tilde \chi^2 + \lambda_1 
      \left( \tilde \phi^2 - \tilde \phi^2_\infty \right)
    \right],
\label{1-80}\\
    \tilde \chi'' + \frac{2}{x} \tilde \chi' &=& \tilde \chi
    \left[
      \tilde \phi^2 + \lambda_2 
      \left( \tilde \chi^2 - \tilde \chi^2_\infty \right)
    \right]
\label{1-90}
\end{eqnarray}
where $x = r \phi(0)$; $\tilde \phi = \phi/\phi(0)$; 
$\tilde \chi = \chi/\phi(0)$; $\tilde \chi_\infty = \chi_\infty/\phi(0)$ and $\tilde \phi_\infty = \phi_\infty/\phi(0)$. Below we will omit $\tilde{}$. The boundary conditions are 
\begin{eqnarray}
  \phi(0) &=& 1, \quad \phi'(0) = 0 ,
\label{1-100}\\
  \chi(0) &=& \chi_0, \quad \chi'(0) = 0 .
\label{1-110}
\end{eqnarray}
The regular solution for equations \eqref{1-80} \eqref{1-90} with boundary conditions \eqref{1-100} \eqref{1-110} does exist for a special choice of 
$\phi_\infty, \chi_\infty$ parameters only. The profiles of $\phi(r), \chi(r)$ in Fig. \ref{functions} are presented. 

The distribution of the gluon condensate 
$\left\langle F^A_{\mu\nu} F^{A\mu\nu} \right\rangle$ in the glueball can be found from the effective Lagrangian \eqref{1-50}. The gluon condensate is
\begin{equation}
	G = - \mathcal L_{eff} =
	\left\langle H^A_i H^{Ai} \right\rangle -
	\left\langle E^A_i E^{Ai} \right\rangle
\label{1-120}
\end{equation}
where $E^A_i, H^A_i$ are chromoelectric and chromomagnetic fields; the signature of 3D metric $\gamma_{ij}$ is positive $\gamma_{ij}=(+,+,+)$; $i,j=1,2,3$ are the space indixes. We see that if $G(x) < 0$ then in this area the chromoelectric field is predominant but if $G(x) > 0$ then the chromomagnetic field is predominant. The substitution of the fields $\phi, \chi$ into the gluon condensate \eqref{1-120} gives us following
\begin{equation}
	G = - \frac{1}{2} {\phi'}^2 - \frac{1}{2} {\chi'}^2 + 
	\frac{\lambda_1}{4} \left(
		\phi^2 - \phi_\infty^2
	\right)^2 + 
	\frac{\lambda_2}{4} \left(
		\chi^2 - \chi_\infty^2
	\right)^2 - \frac{\lambda_2}{4} \chi_\infty^4 - 
	\frac{1}{2} \phi^2 \chi^2 .
\label{1-130}
\end{equation}
The profile of $G(x)$ is presented in Fig. \ref{condensate}. We see that in the glueball the quantum fluctuations of chromoelectric field are predominated. 

\begin{figure}[h]
  \begin{minipage}[t]{.45\linewidth}
  \begin{center}
    \includegraphics[height=1.00\linewidth,width=1.00\linewidth,angle=-90]
    {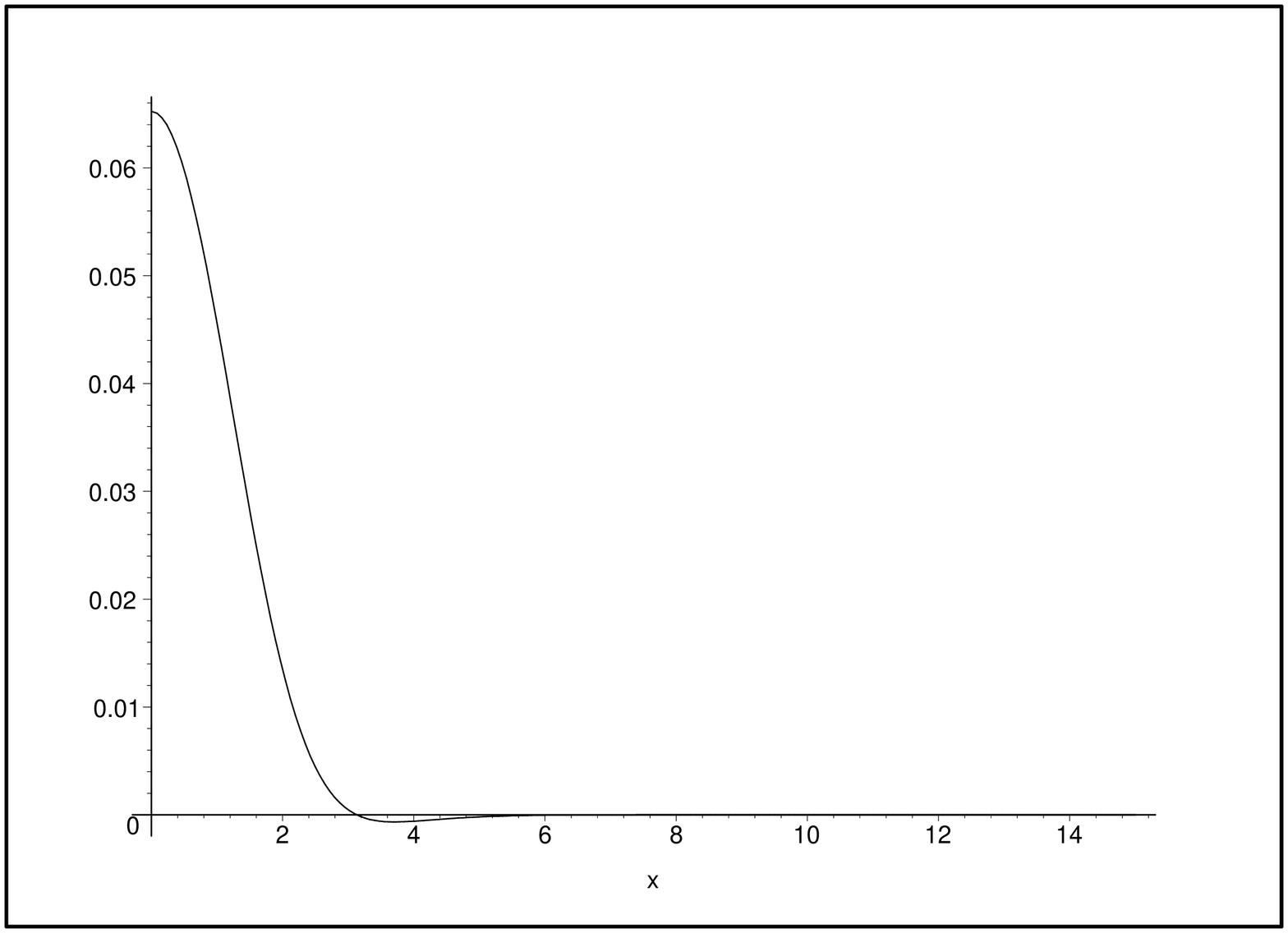}
    \caption{The profile of SU(2) gluon condensate in the glueball.}
    \label{su2condensate}
  \end{center}  
  \end{minipage} 
  \begin{minipage}[t]{.45\linewidth}
  \begin{center}
    \fbox{
    \includegraphics[height=1.00\linewidth,width=1.00\linewidth]{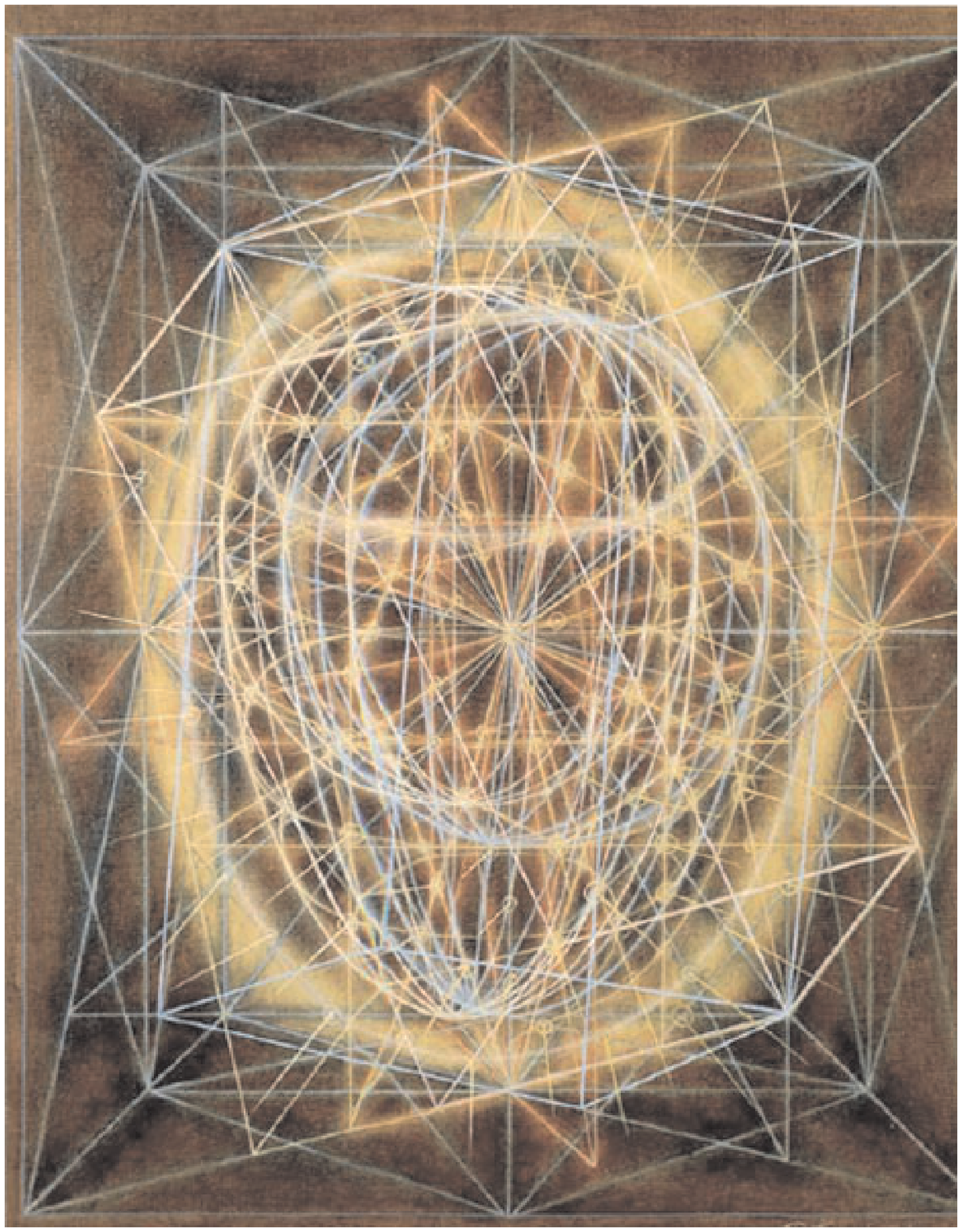}}
    \caption{``Castagna'' of P. Tchelitchew.}
    \label{tchelitchew}
  \end{center}  
  \end{minipage} 
\end{figure}

It is interesting to take a look on the distribution of a SU(2) gluon condensate $G_{SU(2)}$ in this model (here the group SU(2) is considered as the subgroup of SU(3) group). One can show that 
\begin{equation}
	G_{SU(2)} = - \frac{1}{2} {\phi'}^2 +\frac{\lambda_1}{4} \left(
		\phi^2 - \phi_\infty^2
	\right)^2 .
\label{1-140}
\end{equation}
The corresponding profile in Fig. \ref{su2condensate} is presented. We see that in this case the fluctuating chromomagnetic field is predominant. 

\section{Discussion and conclusions}

Thus we have calculated the distribution of the SU(3) gluon condensate in the glueball. In the presented nonperturbative model the glueball is not a cloud of gluon-quanta but this object is more similar to a turbulent liquid with fluctuating velocities and consequently fluctuating lines of fluid flow. In this connection one can say that an object presented in the painting ``Castagna'' of famous surrealist P. Tchelitchew (see Fig. \ref{tchelitchew}) is a schematical presentation of the glueball with fluctuating force lines of chromoelectric fields. 

This result can be compared with the results of Ref. \cite{Dzhunushaliev:2010qs}: the difference is that in the glueball does not exist a core with predominant chromoelectric field in contrast with the flux tube where there does exist a core filled with a longitudinal chromoelectic field stretched between quark and antiquark and confined with a fluctuating  chromomagnetic field. In our opinion the reason for this is that in the core of the flux tube (considered in Ref. \cite{Dzhunushaliev:2010qs}) there exists almost classical chromoelectric field. Whereas in the glueball we have fluctuating chromofields with zero expectation values only. 

\section*{Acknowledgements}

I am grateful to the Research Group Linkage Programme of the Alexander  von
Humboldt Foundation for the support of this research and would like to express
the gratitude to the Department of Physics of the Carl von Ossietzky University
of Oldenburg  and, specially, to J. Kunz.


\begin{thebibliography}{99}

\bibitem{Banks}
T. Banks, R. Horsley, H.R. Rubinstein, and U. Wolff,
Nucl. Phys. \textbf{B190}, 692 (1981);
A.Di Giacomo and G.C. Rossi,
Phys. Lett. B \textbf{100}, 481 (1981);
P.E. Rakow,
PoS LAT2005, 284 (2006) and references therein.

\bibitem{Shifman:1978bx}
  M.~A.~Shifman, A.~I.~Vainshtein, V.~I.~Zakharov,
  Nucl.\ Phys.\  {\bf B147}, 385-447 (1979). 
  
\bibitem{Mathieu:2008me}
  V.~Mathieu, N.~Kochelev, V.~Vento,
  Int.\ J.\ Mod.\ Phys.\  {\bf E18}, 1-49 (2009).
  [arXiv:0810.4453 [hep-ph]]. 
  
\bibitem{Dzhunushaliev:2010qs}
  V.~Dzhunushaliev,
  ``SU(3) flux tube gluon condensate,''
  arXiv:1010.1621 [hep-ph].

\bibitem{Dzhunushaliev:2003sq}
V.~Dzhunushaliev, 
``Scalar model of the glueball,'' 
Hadronic J.\ Suppl.\  {\bf 19}, 185 (2004); 
hep-ph/0312289.

\end{thebibliography}
\end{document}